\documentstyle[preprint,prd,aps]{revtex}

\def\bGam{{\bar\Gamma}}

\def\bbeta{{\mbox{\boldmath $\beta$}}}

\def\gg{{\mathcal{G}}}

\def\Rb{{\bar R}}

\def\d{{\partial}}

\def\dzeroh{{\hat\partial_0}}

\def\Boxh{\hat{\mbox{\kern-.0em\lower.3ex\hbox{$\Box$}}}}

\def\Lie{{\pounds}}

\def\bgrad{{\bar\nabla}}

\newcommand\beq{\begin{equation}}
\newcommand\eeq{\end{equation}}
\newcommand\bea{\begin{eqnarray}}
\newcommand\eea{\end{eqnarray}}
\newcommand\beqa{\begin{eqnarray}}
\newcommand\eeqa{\end{eqnarray}}

\begin{document}
\draft
%

\title{Fixing Einstein's equations 
}

\author{Arlen Anderson and James W. York, Jr.}
\address{
Dept. Physics and Astronomy,
Univ. North Carolina,
Chapel Hill NC 27599-3255}
\date{Jan. 6, 1999}

\maketitle
\vspace{-5.3cm}
\hfill IFP-UNC-528

\vspace{-.3cm}
\hfill gr-qc/9901021
\vspace{4.9cm}

\tightenlines



Einstein's theory of general relativity has not only proven to be physically 
accurate\cite{Wil}, it also sets a standard for mathematical beauty 
and elegance---geometrically.   When viewed as a dynamical
system of equations for evolving initial data, however, 
these equations have a serious flaw:  they cannot be proven to be well-posed
(except in special coordinates\cite{FB52,FiM72,Fri85}). That is,
they do not produce unique solutions that depend smoothly on the initial data.  
To remedy this failing, there has
been widespread interest recently in reformulating Einstein's theory as 
a hyperbolic system of differential equations
\cite{Fri81a,Fri81b,CBR83,Fri91,BM92,FrR94,BMSS95,CBY95,AACbY95,AACbY96,Fri96,FrR96,VPE96,AACbY97,cby97,acby97}.  
The physical and geometrical content of the original theory remain
unchanged, but dynamical evolution is made sound.  
Here we present a new hyperbolic formulation that is strikingly close 
to the space-plus-time (``3+1'') form of Einstein's original equations. 
Indeed, the
familiarity of its constituents make the existence of this formulation
all the more unexpected. This is the most economical
first-order symmetrizable hyperbolic formulation presently known to us 
that has only physical characteristic speeds, either zero or the 
speed of light, for all (non-matter) variables.
It also serves as a foundation for unifying previous proposals. 

The source of the imperfection in Einstein's theory lies in the fact that 
it is a constrained theory.  Physical initial data cannot be freely specified, 
yet even infinitesimally perturbed data that violate the physical constraints 
can lead to results so wildly divergent that they spoil the desired smooth 
dependence on initial data.  This is particularly 
troublesome in numerical evolution where such violations are unavoidable.
The lack of well-posedness is also a serious problem when addressing such a 
basic question as the global nonlinear stability of flat Minkowski spacetime.
In fact, the proof of such stability\cite{ChK} employs the hyperbolic
wave equation of \cite{CBR83} which we discuss below.  A well-posed
formulation of Einstein's equations would also seem to be an essential 
starting point for the conventional approach to quantum gravity in which 
one first quantizes the (unconstrained) classical theory and then imposes the 
constraints.

The desire to simulate numerically the full nonlinear evolution of 
Einstein's equations 
in three dimensions, such as in the collision of two black holes 
({\it cf., e.g.,} \cite{abr98}), has motivated much of the recent
effort on hyperbolic formulations:  well-posed underlying equations 
make stable numerical evolution much more likely than otherwise 
would be the case, and  formulations cast in
first-order symmetrizable form are especially suited to
numerical implementation.  In addition, physical characteristic speeds make 
it easy to impose good boundary conditions, crucial
to a successful numerical scheme.  To amplify this point,
Einstein's equations contain many unphysical (``gauge'') variables
among its unknowns, 
and {\it a priori} they can travel at any speed.  A formulation with only
physical characteristic speeds has significant advantages because no
explicit separation of physical and unphysical degrees of freedom is
required.  The physical and unphysical variables propagate at the same speeds 
and therefore satisfy boundary conditions on the same characteristic
surfaces.  This is particularly important, for example, at the horizon of a
black hole, which is a characteristic boundary for physical variables but not 
for unphysical ones, unless the latter propagate at the speed of light. 

The following system of thirty equations will be shown to be 
symmetrizable hyperbolic 
\beq
\label{dg}
0 = \dzeroh g_{ij} + 2N K_{ij},
\eeq
\beq
\label{Rij}
R_{ij} = -N^{-1} \dzeroh K_{ij} + \Rb^{(e)}_{ij} 
- N^{-1} \bgrad_i \bgrad_j N(\alpha,g)
+ K K_{ij} - 2 K_{ik} K^k\mathstrut_j, 
\eeq
\beq
\label{R0k_gam}
2 g_{ij} R_{k0} = \dzeroh \bGam_{kij} + \d_j(N K_{ki}) +
\d_i (N K_{kj}) - \d_k (N K_{ij})
-2 g_{ij} N \bgrad^m (K_{km} - g_{km} K)
\eeq
(notation elaborated below).  This form suggests the
name ``Einstein-Christoffel system.'' It is convenient for pedagogical
reasons to replace the third equation by the equivalent equation
\beq
\label{R0k_gg}
4 g_{k(i} R_{j)0} =  \dzeroh \gg_{kij} + \d_k (2N K_{ij}) 
-4g_{k(i} N\bgrad^m(K_{j)m} - g_{j)m}K)
\eeq
The indices $i$, $j$, $k$, $m$ run over the spatial indices 1, 2, 3, and
the Einstein convention of summing over repeated indices 
is used. A short-hand notation to indicate symmetrization, 
$A_{(i}B_{j)}=(1/2)(A_i B_j + A_j B_i)$, is used in the previous equation. 

To establish notation, assume that spacetime has topology
$\Sigma \times R$ with metric given in the foliation-adapted basis,
\beq
\label{metric}
ds^2 = -N^2 (dt)^2 + g_{ij} (dx^i +\beta^i dt)(dx^j + \beta^j dt).
\eeq
Here, $N(\alpha,g)$ is the lapse scalar, and $\beta^i(x,t)$ is 
the spatial shift vector, freely specifiable on the spacelike slices 
$t={\rm constant}$. 
The lapse $N$ is determined through $N=\alpha g^{1/2}$, where
$\alpha(x,t)$ is a freely specified ``slicing'' density (of weight $-1$) and
$g=\det g_{ij}$ is the determinant of the spatial metric $g_{ij}$.
The spatial derivatives of the metric are denoted by 
\beq
\label{gg_def}
\gg_{kij} = \d_k g_{ij}.
\eeq
This is a subtle element, as it will transpire that while this relation
is imposed initially, it may not hold for the evolved quantities,
as discussed below.
The spatial Christoffel symbols in this metric are,
with $\bGam^m\mathstrut_{ij}=g^{mk}\bGam_{kij}$,
\beq
\label{bGam_def}
\bGam_{kij}(\gg)\equiv (1/2)(\gg_{jki} + \gg_{ikj} -\gg_{kij}).
\eeq
To focus attention on $\gg_{kij}$, the Christoffel symbols will not be 
used here as independent variables, though they could be, but only as 
a compact notation for 
this expression in terms of $\gg_{kij}$; the explicit 
functional dependence ``($\gg$)'' is intended to reflect this choice.  
Finally, $K_{ij}$ denotes the extrinsic curvature of the slice $\Sigma$, and
$K=K^k\mathstrut_k$ is its trace.

The derivative $\bgrad_k$ is the spatial covariant derivative operator 
in $\Sigma$.  The derivative $\dzeroh=\d_t -
\Lie_\bbeta$, where $\d_t=\d/\d t$ and $\Lie_\bbeta$ is the Lie derivative 
along the
shift vector $\bbeta$ in a $t={\rm constant}$ slice, is the natural time
derivative for evolving time-dependent spatial tensors.  It is the
extension to tensors of the (non-coordinate) basis vector $\d_0=\d_t
-\beta^k \d_k$ ($\d_k=\d/\d x^k$) that is normal to the slice $\Sigma$.  
Note that while
$[\d_0,\d_j] =\d_0 \d_j - \d_j \d_0=(\d_j \beta^k)\d_k \ne 0$,
\beq
\label{dzeroh_comm}
[\dzeroh,\d_k]=0.
\eeq

On the left-hand side of (\ref{Rij}) and (\ref{R0k_gam}) or
(\ref{R0k_gg}), $R_{ij}$ and $R_{j0}$
are spacetime Ricci curvature tensors and are to be replaced by
their appropriate expressions
in terms of matter stress tensors from Einstein's equations,
$G_{\mu\nu}\equiv R_{\mu\nu} - (1/2) g_{\mu\nu} R = 8\pi T_{\mu\nu}$.
Here, $\mu$, $\nu$ run from 0 to 3, $R=R^\mu\mathstrut_\mu$, and $T_{\mu\nu}$
is the matter stress tensor.   $\Rb^{(e)}_{ij}$ is the spatial Ricci 
curvature
tensor of the spacelike slice $\Sigma$.  It is essential to manipulate
the standard form of $\Rb_{ij} = \d_k \bGam^k\mathstrut_{ij} 
- \d_j \bGam^k\mathstrut_{ik} + \bGam^k\mathstrut_{mk} \bGam^m\mathstrut_{ij}
- \bGam^k\mathstrut_{mj} \bGam^m\mathstrut_{ik}$ into a distinct 
but (initially) equivalent form, indicated below, and the 
superscript ``$(e)$'' reflects this change.

Einsteinian initial data for the system (\ref{dg}), (\ref{Rij}), (\ref{R0k_gg})
are $g_{ij}$, $K_{ij}$, $\gg_{kij}$, specified on an initial slice 
$\Sigma_0$, and presumed to satisfy the Einstein constraints,
$G^0\mathstrut_0=8\pi T^0\mathstrut_0$ and 
$R^{0}\mathstrut_k=8\pi T^0\mathstrut_k$.  This system of
initial constraints is well understood as a semilinear elliptic
system\cite{Yor79,CBY80}.  A mathematically well-posed form of
the twice-contracted Bianchi identities\cite{Fri97,ay98} shows that
these initial-value constraints remain satisfied if the equations of motion 
are equivalent to 
$R_{ij}=8\pi(T_{ij}-(1/2)g_{ij}T^\mu\mathstrut_\mu)$.

That the system (\ref{dg}), (\ref{Rij}), (\ref{R0k_gg}) is hyperbolic is 
not obvious, but its content is easy to grasp.  The first equation (\ref{dg}) 
is simply a definition of the extrinsic curvature $K_{ij}$.  The second 
equation (\ref{Rij}) is the 3+1 decomposition of the space-space components 
of the spacetime Ricci tensor.  As such, these are the basic geometric
ingredients of Einstein's original equations and of all 3+1 formulations 
of general relativity \cite{CB56,Yor79,CBY80}.  The
remarkable fact is that the equation (\ref{R0k_gg}) completes the
first two into a symmetrizable hyperbolic system.  The content of
equation (\ref{R0k_gg})  is also readily understood.  

If one applies $\dzeroh$ to (\ref{gg_def}) and uses (\ref{dzeroh_comm}) 
and (\ref{dg}), one obtains the identity
\beq
\label{dgg_id}
\dzeroh \gg_{kij} = -\d_k(2N K_{ij}).
\eeq
This is the
right hand side of (\ref{R0k_gg}) aside from the 3+1 decomposition of
$4g_{k(i}R_{j)0}$.  Ordinarily in Einstein's theory, 
\beq
\label{R0k_def}
R_{0j} =-N\bgrad^m(K_{jm}-g_{jm} K)
\eeq
is a constraint---the ``momentum''
constraint---because it involves no time derivatives.  What is special about 
(\ref{R0k_gg}) is that it makes the momentum constraint  dynamical by 
combining it with the identity (\ref{dgg_id}) involving a time-derivative.  
This defines a modified evolution of $\gg_{kij}$ when the constraint
is not satisfied, that is, when (\ref{R0k_def}) does not hold after $R_{0j}$
is replaced by its matter expression.  

The identity (\ref{dgg_id}) is closely related to metric compatibility
of the connection.   In a general spatial frame, 
a connection is metric compatible if and only if 
$\bGam_{ijk} + \bGam_{jik}=\d_k g_{ij}$.
Taking the time derivative of this condition and applying (\ref{R0k_gg}) gives
\beq
\label{bGam_evol}
2\dzeroh \bGam_{(ij)k}= - \d_k (2 N K_{ij}) 
+4 g_{k(i} N \bgrad^m (K_{j)m} - g_{j)m} K) + 4 g_{k(i}R_{j)0}.
\eeq
Thus, if the momentum constraint were violated, metric compatibility of
$\bGam$ would be lost in the evolution of (\ref{bGam_evol}). 
While $2\bGam_{(ij)k}=\gg_{kij}$ always holds, the evolved $\gg_{kij}$
would no longer be the spatial derivative $\d_k$ of the evolved $g_{ij}$.

To motivate the system (\ref{dg}), (\ref{Rij}), (\ref{R0k_gg}) further, 
consider two of its predecessors, the Einstein-Ricci
formulation \cite{CBR83,CBY95,AACbY95,AACbY96,AACbY97} and the Frittelli-Reula 
formulation \cite{FrR94,FrR96}.
The third-order Einstein-Ricci system consists of (\ref{dg})
and a wave equation built from
(\ref{Rij}) and (\ref{R0k_def}) through the combination
\beq
\label{ER}
\dzeroh R_{ij} - \bgrad_i R_{j0} - \bgrad_j R_{i0} = N \Boxh K_{ij}
+J_{ij} + S_{ij}.
\eeq
It is called third-order because of the effective number of derivatives
of $g_{ij}$ in (\ref{ER}).
Here, $\Boxh = -N^{-1} \dzeroh N^{-1} \dzeroh + \bgrad^k \bgrad_k$.
$J_{ij}$ is a nonlinear function of $K_{ij}$, $N$, their first derivatives,
and the second derivatives of $N$.  $S_{ij}$ is a potentially troublesome 
term involving a second spatial derivative of $K$ and a third derivative of 
$N$.  The behavior of $S_{ij}$ is tamed by using
$N=\alpha(x,t) g^{1/2}$ \cite{CBR83,CBY95} (or by imposing generalized 
harmonic slicing \cite{AACbY97} with a gauge source \cite{Fri85}).  Note 
that the use of $\alpha$ permits any time-slicing to be employed.  

This system can be put in first-order form by
introducing new variables to represent the temporal and spatial derivatives
of $K_{ij}$ and of $N$.  Together with $(\ref{dg})$ and the equations obtained
by applying $\dzeroh$ to $\bGam^{k}\mathstrut_{ij}$, and using
$\alpha$ to eliminate $N$, a system of 66 equations is found.
This system is manifestly spatially covariant, is expressed in 3+1 geometric 
variables, and has only physical characteristic speeds, either zero or 
the speed of light.  

One may wonder about the large number
of equations and about a deeper meaning behind the
combination in (\ref{ER}).  Regarding the number of equations, the 
Einstein-Ricci system is equivalent (for Einsteinian initial data) to the 
Einstein-Bianchi system\cite{cby97,acby97} which also has 66 equations.  
There, it is evident that this number of equations is precisely that needed 
to incorporate the full Bianchi identities and to propagate the Riemann 
curvature tensor explicitly in a system having only physical characteristics
(otherwise, {\it cf.} \cite{Fri96}).  

The Frittelli-Reula system\cite{FrR94,FrR96}, in contrast, has 30 equations, is 
expressed in 
non-covariant variables, and admits superluminal characteristic speeds for
some (unphysical) degrees of freedom.  It was constructed by
proposing an energy norm built from the extrinsic curvature and
spatial derivatives of the metric.  Frittelli and Reula parametrize
their construction and find a large family of hyperbolic systems with
different characteristics, none wholly physical.  Friedrich has
observed that an equation for the metric constructed from (\ref{dg}) 
and (\ref{Rij}), while not of known hyperbolic type, has only physical 
characteristics \cite{Fri96}.   A natural question is 
whether there are further thirty-variable hyperbolic systems 
and whether any has only physical characteristics.
The Einstein-Christoffel system presented here is such a system, and 
it is easy from it to see how to extend the Frittelli-Reula construction. 

The Einstein-Ricci system has only physical characteristics, but its equations
number more than twice those of the Frittelli-Reula system.  A natural
question is whether the third-order form can be put in first-order
form to achieve a thirty variable system.  The answer is yes.  To see
why this might be possible, consider the simple wave equation
\beq
\label{wave_eq}
\d_t^2 u - \d_x^2 u =0,
\eeq
This can be put in first order form in two ways.  The easiest
is to introduce the derivatives of the dependent variable as new
variables.  Introduce $U=\d_t u$ and $V=\d_x u$ to reach the system
\beqa
\d_t u &=& U, \\
\d_t U -\d_x V &=& 0, \nonumber \\
\d_t V - \d_x U &=& 0. \nonumber
\eeqa
The last equation is an integrability condition reflecting the
commutativity of the partial derivatives.  This parallels the way
that the first-order form of the Einstein-Ricci system was obtained
from (\ref{ER}). 

The second way to get to first order form is to pull apart the wave
equation to obtain first order pieces
\beqa
\label{fo2}
\d_t u - \d_x v &=& 0, \\
\d_t v - \d_x u &=& 0. \nonumber
\eeqa
The first method is essentially one derivative higher.  It necessarily 
involves more variables.  Note that the wave equation (\ref{wave_eq})
is reconstructed from this system (\ref{fo2}) by taking a time derivative of 
the first equation and adding a spatial derivative of the second.  This
parallels the structure in (\ref{ER}) that leads to the third-order 
Einstein-Ricci system.  This encourages the speculation that a
``pulled-apart'' system analogous to (\ref{fo2}) is possible.  The 
obstacle however is that the momentum constraint
as usually construed is not a dynamical equation, so the obvious
``pulled-apart'' system is not a hyperbolic system.    The key 
idea is that adding a suitably chosen dynamical identity to the 
momentum constraint overcomes this obstacle and leads 
to a symmetrizable hyperbolic system.

To begin, we work with $\gg_{kij}$ rather than 
$\bGam_{kij}$. Focus on the derivatives of the Christoffel symbols 
contained in $\Rb_{ij} -N^{-1}\bgrad_i\bgrad_j N$.  These are the
essential terms from the standpoint of hyperbolicity.  [Recall
that $N=\alpha g^{1/2}$, so $\bgrad_j N = g^{1/2} \d_j \alpha
+ \bGam^k\mathstrut_{jk}(\gg) g^{1/2} \alpha$.]  These terms are 
reorganized as follows:
\beqa
\d_k \bGam^k\mathstrut_{ij}(\gg) - \d_j \bGam^k\mathstrut_{ik}(\gg) 
- \d_i \bGam^k\mathstrut_{jk}(\gg)
&=& \\
&& \hspace{-2.5cm} =
{1\over 2} \biggl( \d_k [g^{km} (\gg_{jmi} + \gg_{imj} - \gg_{mij})] -
\d_j( g^{rs} \gg_{irs}) - \d_i ( g^{rs} \gg_{jrs} )\biggr) \nonumber \\
&& \hspace{-2.5cm} ={1\over 2} \biggl( -\d_k (g^{km} \gg_{mij})
+ \d_i [g^{rs}(\gg_{rsj} - \gg_{jrs})]
+ \d_j [g^{rs}(\gg_{rsi} - \gg_{irs})] \nonumber \\
&&\hspace{-1.5cm} +g^{kr} g^{sm} (\gg_{irs} \gg_{kmj}
+ \gg_{jrs} \gg_{kmi}) - g^{kr} g^{sm} \gg_{krs} (\gg_{jmi} + \gg_{imj})
\biggr),
\nonumber
\eeqa
where $\d_k (g^{km} \gg_{jmi}) = \d_j (g^{km} \gg_{kmi}) 
+g^{kr} g^{sm} (\gg_{jrs} \gg_{kmi}-\gg_{krs} \gg_{jmi})$ 
has been used.
Introducing
\beqa
\label{f_def}
f_{kij} &\equiv& {1\over 2}[ \gg_{kij} - g_{ki} g^{rs} (\gg_{rsj} - \gg_{jrs})
-g_{kj} g^{rs} (\gg_{rsi} - \gg_{irs})] \\
&=& \bGam_{(ij)k} + {1\over 2} g_{ki} g^{rs} (\bGam_{rsj} - \bGam_{jrs})
+{1\over 2} g_{kj} g^{rs} (\bGam_{rsi} - \bGam_{irs}) \nonumber 
\eeqa
puts the leading derivatives of (\ref{Rij}) in the familiar form
\beq
\label{Rij3}
R_{ij} = -N^{-1} \dzeroh K_{ij} - \d^k f_{kij}+ l.o._{ij},
\eeq
where $l.o._{ij}$ stands for lower order terms containing no derivatives of
unknowns. They are
\beqa
l.o._{ij} &=& H K_{ij} - 2 K_{ik} K^k\mathstrut_j
- \alpha^{-1} \bgrad_i \d_j \alpha
-  \bGam^k\mathstrut_{jk}(\gg) \alpha^{-1} \bgrad_i \alpha \\
&& + 2 \bGam^k\mathstrut_{mk}(\gg) \bGam^m\mathstrut_{ij}(\gg) -
\bGam^k\mathstrut_{mj}(\gg) \bGam^m\mathstrut_{ik}(\gg)
-g^{kr} g^{sm} \gg_{krs} f_{mij} \nonumber \\
&& +{1\over 2} [ g^{kr} g^{sm}( \gg_{jrs} \gg_{kmi} + \gg_{irs} \gg_{kmj})
-g^{kr} g^{sm} \gg_{krs}( \gg_{jmi} + \gg_{imj} )]. \nonumber
\eeqa

Turn to consider (\ref{R0k_gg}). From (\ref{f_def}), one computes
\beq
\label{R0j3}
g_{ki} R_{j0} + g_{kj} R_{i0} = - \dzeroh f_{kij} - \d_k (N K_{ij})
+ l.o._{kij}.
\eeq
The lower order terms are 
\beqa
\label{lokij}
l.o._{kij}&=& N K_{ki} g^{rs} ( \gg_{rsj} - \gg_{jrs})
+N K_{kj} g^{rs} ( \gg_{rsi} - \gg_{irs}) \\
&&+ g_{ki}[K_{mj} \d^m N -H \d_j N + N g^{rs} \bGam^k\mathstrut_{jr}(\gg) K_{sk
}
-N (\gg_{rsj} - 2 \gg_{jrs}) K^{rs}] \nonumber \\
&&+ g_{kj}[K_{mi} \d^m N -H \d_i N + N g^{rs} \bGam^k\mathstrut_{ir}(\gg) K_{sk
}
-N (\gg_{rsi} - 2 \gg_{irs}) K^{rs}]. \nonumber
\eeqa

The subsystem (\ref{Rij3}) and (\ref{R0j3}) [completed by (\ref{dg})] is 
obviously symmetrizable hyperbolic because it has the familiar structure
of a wave equation in first order form.  [It is not rigorously symmetric 
hyperbolic because
a metric is present to raise the spatial derivative index in (\ref{Rij3}).]
It is also clear that to build a wave equation in $K_{ij}$
from (\ref{Rij3}) and (\ref{R0j3}), one forms the combination
in (\ref{ER}).  This reveals the meaning behind this combination.  
The characteristic speed in the subsystem (\ref{Rij3}), (\ref{R0j3}) is the 
speed of light,
so the extrinsic curvature and the connection propagate at the speed
of light.  From (\ref{dg}), the metric propagates at speed zero. 

It should be emphasized that the wave equation obtained 
from (\ref{Rij3}), (\ref{R0j3}) is not
exactly the same as the third-order Einstein-Ricci system.  They
differ by lower order terms proportional to constraints.  This means
that they will agree for Einsteinian initial data, but may disagree
when the constraints are violated.

The energy norm for the system (\ref{dg}), (\ref{Rij3}), (\ref{R0j3}) is 
the integral over $\Sigma$ of $K^{ij} K_{ij} + f^{kij} f_{kij}$, 
where $f^{kij} = g^{km} g^{ir} g^{js} f_{mrs}$. 
When the energy norm is expressed in terms of $P_{ij} = K_{ij} - g_{ij} K$, 
$P=P^k\mathstrut_k$, 
and $M_{kij}=-(1/2)(\gg_{kij} -g_{ij} g^{rs} \gg_{krs})$, the result
can be compared to the {\it ansatz} of Frittelli-Reula \cite{FrR96}.  
Additional terms are present beyond those that they considered.  
Their energy norm {\it ansatz} can be generalized by adding arbitrary positive 
multiples of quadratic terms formed from traces of $P_{ij}$ and of $M_{ijk}$. 
Similarly, their parametrization of $M_{ijk}$ can be extended using 
combinations of traces of $\gg_{ijk}$, such as occur in $f_{kij}$. This 
produces a larger many-parameter family of symmetrizable hyperbolic systems 
equivalent to Einstein's equations. Some of these other systems also have only
physical characteristics, for example, a multiple of the Hamiltonian
constraint $G^0\mathstrut_0$ can be added to (\ref{Rij}) if a
compensating multiple of the momentum constraint $R_{0k}$ is added to 
(\ref{R0k_gam}) and (\ref{R0k_gg}). 

The Einstein-Christoffel system discussed here is a well-posed system of 
30 equations that has only physical characteristic speeds and can be expressed 
in 3+1 geometric variables.  Spatial covariance of the system is not explicit,
but present nonetheless.  This formulation clarifies the relationships among 
Einstein's original equations and the Einstein-Ricci and 
Frittelli-Reula hyperbolic formulations.  One can see further links 
to the Einstein-Bianchi, Friedrich \cite{Fri81a,Fri81b,Fri91,Fri96}
and Bona-Mass\'o \cite{BM92,BMSS95} formulations.
It will be very interesting to implement
this system numerically to see how it behaves.  Having only
physical characteristics should prove very useful when imposing
boundary conditions on the horizon of a black hole since no
information, physical or otherwise, can leave the black hole,
and all information will enter the black hole in a physical
way. 
 
Acknowledgement.  AA thanks Mark Hannam for discussions. 
This work was supported in part by the NSF.

\end{document}